\documentclass[journal,a4paper,10pt]{IEEEtran}
\usepackage{cite}
\usepackage{flushend}
\usepackage{graphicx,psfrag}

 \usepackage[cmex10]{amsmath}
 \usepackage{amsfonts}
 \usepackage{etex}
\usepackage{pgfplots}
\usepackage{tikz}

\usepackage{marginnote} 

\usepackage{marginnote}
\graphicspath{{figs/}}

\author{Carl Gustafson, Taimoor Abbas, David Bolin and Fredrik Tufvesson%
\thanks{This work was supported by the VINNOVA FFI program Wireless Communications in Automitve Environments.}%
\thanks{C. Gustafson and F. Tufvesson are with the Dept. of Electrical and Information Technology, Lund University, Sweden.}%
\thanks{T. Abbas is with the Volvo Car Corporation, Gothenburg, Sweden.}%
\thanks{D. Bolin is with the Mathematical Sciences, Chalmers University of Technology, Gothenburg, Sweden.}}
\title{Statistical Modeling and Estimation of Censored Pathloss Data}

\begin{document}
\maketitle

\begin{abstract}
Pathloss is typically modeled using a log-distance power law with a large-scale fading term that is log-normal. However, the received signal is affected by the dynamic range and noise floor of the measurement system used to sound the channel, which can cause measurement samples to be truncated or censored. If the information about the censored samples are not included in the estimation method, as in ordinary least squares estimation, it can result in biased estimation of both the pathloss exponent and the large scale fading. This can be solved by applying a Tobit maximum-likelihood estimator, which provides consistent estimates for the pathloss parameters. This letter provides information about the Tobit maximum-likelihood estimator and its asymptotic variance under certain conditions. 
\end{abstract}
\begin{IEEEkeywords}
Pathloss, maximum-likelihood estimation, ordinary least squares, censored data, truncated data, vehicular communication. 
\end{IEEEkeywords}

\section{Introduction}
Pathloss describes the expected loss in received power as a function of the transmitter (Tx) and receiver (Rx) separation distance and the effects of random large scale fading. It includes losses due to the expansion of the radio wave front in space as well as losses due to reflection, scattering, diffraction and penetration. A number of pathloss models have been developed for a variety of wireless communication systems, e.g., for cellular systems, Bluetooth, Wi-Fi, vehicle-to-vehicle communications, and, mm-wave point-to-point communications, operating over different frequency bands ranging from hundreds of MHz to tens of GHz \cite{Hatay1980, Erceg1999, Guoqiang2006, molisch09-CommMag}. These models have widely been used for the prediction and simulation of signal strengths for given Tx-Rx separation distances. Pathloss models are often developed based on channel measurements in realistic user scenarios. The model parameters estimated from measurement data are thus typically valid only for a particular frequency range, antenna arrangement, and environment for the targeted user scenario. 

However, in practice, the observation of the received signal power at the receiver is limited by the system noise, i.e., the signals with power below the noise floor can not be measured properly. In many vehicle-to-vehicle measurements, this limitation due to the system noise is often present at longer distances \cite{Kunisch,Mahler,Onubogu}. Also, in mm-wave measurements, the pathloss values are in general larger than at lower frequencies, which effectively can reduce the range in which the data is unaffected by the noise floor. Due to the limited dynamic range of the measurement system, sample data might be \emph{truncated}, whereby all data above or below a certain range are immeasurable, or \emph{censored}, meaning that all data above or below a certain range are counted, but not measured. Estimation of the cluster decay and cluster fading based on truncated data has previously been addressed in \cite{Cluster}. For clusters, the data is modeled as truncated, since it is generally impossible to measure or count clusters that are below the noise floor. However, in pathloss measurements, where distances for the measurement points where the  received power falls below the noise floor are known, it is possible to model the data as being censored.   Estimating statistical parameters without considering the effects of censored or truncated data samples, can lead to erroneous results. The fact that this can be a problem for pathloss data is acknowledged in \cite{Mahler}, however, the authors do not give any detailed information on how to solve this issue. 
In this letter, we discuss the use of a Tobit model \cite{Tobin} for censored pathloss data and a maximum-likelihood (ML) method for the estimation of pathloss parameters \cite{Takeshi}. Supplementary material and Matlab codes can be found in a supporting technical report \cite{Technical}.  

\section{Pathloss Modeling}
Pathloss is often modeled by a log-distance power law plus a large scale fading term \cite{Molisch}. In units of dB this can be written as 
\begin{eqnarray}
PL(d)=PL(d_{0})+10n\mathrm{log}_{10}\left(\frac{d}{d_{0}}\right)+\Psi_{\sigma}, \ \ d\geq d_{0}, 
\label{eq:PowerLaw}
\end{eqnarray}
 where $d$ is the distance, $n$ is the pathloss exponent, $PL(d_{0})$ is the pathloss at a reference distance of $d_{0}$ and $\Psi_{\sigma}$ is a random variable that describes the large-scale fading around the distance-dependent mean pathloss. For measurement data, it is here assumed that the effects of small scale fading is removed from the data set. It is also assumed that the peak value of the aggregated antenna gain is removed from the measurement data\cite{Johan}. Ideally, the variation of the aggregated antenna gain should be small, so that it does not affect the measured large-scale fading too much. The large-scale fading term is usually modeled by a log-normal distribution, which in the dB-domain corresponds to a zero-mean Gaussian distribution with standard deviation $\sigma$, i.e., $\Psi_\sigma \sim \mathcal{N}(0,\sigma^2)$. Hence, the pathloss is normally distributed with a distance dependent expected value, $PL(d)\sim \mathcal{N}(\mu(d),\sigma^2)$, where

\begin{equation}
\mu(d) = PL(d_{0})+10n\mathrm{log}_{10}\left(\frac{d}{d_0}\right).
\label{eq:det_mean}
\end{equation}
 
The reference value $PL(d_{0})$ can be estimated based on measurement data, or based on reference measurements at this distance. For line-of-sight (LOS) scenarios, it is sometimes deterministically modeled based on the free-space pathloss, as
 \begin{eqnarray}
PL(d_{0})=20\mathrm{log}_{10}\left(\frac{4\pi d_{0}}{\lambda}\right). \label{FSPLref}
\end{eqnarray}
Here $\lambda$ is the wavelength at the given frequency. Here, it is worth noting that the approach of using the deterministic reference value of Eq.~(\ref{FSPLref}) only provides theoretically correct results if the pathloss exponent is equal to 2. If the pathloss exponent is \emph{not} equal to 2, but Eq.~(\ref{FSPLref}) is used to determine the reference value, the data model of Eq.~\ref{eq:PowerLaw} is inconsistent, as it depends on the choice of the reference distance $d_{0}$. For non-line-of-sight (NLOS) scenarios, it is clear that the free-space equation (\ref{FSPLref}) does not hold, which means that the reference value in this case must be determined in another fashion. Due to the above, it is preferable to use actual measurements of the reference level, or, to  estimate it based on the measurement data. In some cases, it might be difficult to produce reliable measurements of the reference value scenarios due to practical reasons, especially considering that it might be hard to produce a large number of uncorrelated measurement samples exactly at $d_{0}$.  

\section{Estimation by Ordinary Least Squares}
To completely model the pathloss and large-scale fading for a given data set, we wish to estimate the three parameters of (\ref{eq:PowerLaw}), i.e., $n$, $PL(d_{0})$ and $\sigma^2$. The data under consideration is implicitly assumed to be Gaussian since $\Psi_{\sigma}$ is Gaussian in the dB domain. Using (\ref{eq:PowerLaw}) the data set for $L$ path loss measurements, $\mathbf{y}=\left[PL(\mathbf{d}/d_{0})\right]_{L\times 1}$ can be written as,

\begin{equation}
\mathbf{y}=\mathbf{X}\boldsymbol\alpha+\boldsymbol\epsilon,
\label{model_1}
\end{equation}
where $\mathbf{X}=[\mathbf{1} \ \ 10\mathrm{log}_{10}(\mathbf{d}/d_{0})]_{L\times 2}$ and $\boldsymbol\alpha=[PL(d_{0}) \ n]^{T}$.
The term $\boldsymbol\epsilon=[\boldsymbol\Psi_{\sigma}]_{L\times 1}$ is a row vector describing the large-scale fading term for each of the $L$ different pathloss samples. 

When there are no censored samples, the parameters of the log-distance power law can be estimated by applying ordinary least squares (OLS). The parameter $ \boldsymbol\alpha$  is then estimated as\footnote{As the variance  $\sigma^{2}$ is assumed to be independent of delay, weighted least squares (WLS) are not applied. However, we note that WLS could be of use for cases when $\sigma^2$ is being modeled with a distance dependence.}

\begin{eqnarray}
\hat{\boldsymbol\alpha}=\left(\mathbf{X}^T\mathbf{X}\right)^{-1}\mathbf{X}^T\mathbf{y}.
\label{eq:ordLS1}
\end{eqnarray}
The variance of the large-scale fading, $\sigma^2$, can then be estimated as
\begin{eqnarray}
\hat{\sigma}^2 = \frac{1}{L-1}(\mathbf{y}-\mathbf{X}\hat{\boldsymbol\alpha})^T(\mathbf{y}-\mathbf{X}\hat{\boldsymbol\alpha}).
\label{eq:ordLS2}
\end{eqnarray}

The estimate $\hat{\boldsymbol\alpha}$ is Gaussian,
\begin{eqnarray}
\hat{\alpha}_{j}\sim\mathcal{N}\left(\alpha_{j},\sigma^2(\mathbf{X}^{T}\mathbf{X})_{jj}^{-1}\right), \ j=1,2,
\end{eqnarray}
which alternatively can be expressed as
\begin{eqnarray}
\begin{aligned}
\hat{PL}(d_{0})&=\hat{\alpha}_{1}\sim\mathcal{N}\left(PL(d_{0}),\sigma^2\left(L^{-1}+\bar{\mathbf{x}}^2S_{xx}^{-1}\right)\right), \\
\hat{n}&=\hat{\alpha}_{2}\sim\mathcal{N}\left(n,\sigma^2S_{xx}^{-1}\right),
\end{aligned}
\end{eqnarray}
where 
\begin{eqnarray}
\begin{aligned}
\bar{\mathbf{x}}&=\frac{1}{L}\sum_{l=1}^{L}10\mathrm{log}_{10}(d_{l}/d_{0}), \\
S_{xx}&=\sum_{l=1}^{L}\left(10\mathrm{log}_{10}(d_{l}/d_{0})-\bar{\mathbf{x}}\right)^2.
\end{aligned}
\end{eqnarray}
Using Eq.~(9), standard errors\footnote{The standard error is the standard deviation of the sampling distribution of a statistic.} for $\hat{n}$ and $\hat{PL}(d_{0})$  can be found by replacing the unknown standard deviation of the large scale fading, $\sigma$, by its estimate, $\hat{\sigma}$, which gives
\begin{eqnarray}
\begin{aligned}
\hat{\mathrm{SE}}(\hat{n})&=\hat{\sigma}\sqrt{S_{xx}^{-1}}, \\
\hat{\mathrm{SE}}(\hat{PL}(d_{0}))&=\hat{\sigma}\sqrt{L^{-1}+\bar{\mathbf{x}}^2S_{xx}^{-1}}. \\
\end{aligned}
\end{eqnarray}
The standard errors are useful for evaluating the accuracy of the estimated parameters. However, it should be stressed that these standard errors only applies when the data actually follows the log-distance power law model with a large-scale fading variance that is independent of delay. For this reason, it is often necessary to validate the measurement data against the presumed model. This could be done by investigating the residuals between the measured data and the regression, to make sure that the residuals do not exhibit any sort of distance dependence. If the data seems to be described by a different model, then a different pathloss model would have to be considered. The standard error of the parameters estimated with OLS depend on the number of samples and the exact pathloss sample distances that are used in the measurement.  However, if the data is being censored, OLS would provide biased results, which means that Eq.~(11) no longer applies.  

\section{Estimation of Censored Pathloss Data}
In order to estimate the pathloss exponent and fading variance of censored data, with a $\emph{known}$ number of missing samples where only the distance is known, it is possible to base the estimation on a censored normal distribution. Under this assumption, the observations follow a censored normal distribution \cite{Tobin}. The censoring occurs for data samples where the pathloss is greater than or equal to $c$. The value $-c$ is a channel gain that corresponds to the noise floor of the channel sounder or measurement device. In practice, $c$ is chosen with some margin with respect to the noise floor, so that a limited number of samples dominated by noise are included as measurement data.
Using the data set model in (\ref{model_1}), the data is assumed to be censored so that observations with values at or above $c$ are set to $c$, i.e.,

\begin{eqnarray}
y_{i} = \left\{ \begin{matrix}
  y_{i}^{*} & \mathrm{if} \ y_{i}^*< c \\
  c & \mathrm{if} \ y_{i}^*\geq c 
 \end{matrix}\right.\,
\end{eqnarray}
where
\begin{eqnarray}
y^{*}_{i}\sim \mathcal{N}(\mathbf{x}_{i}\boldsymbol\alpha,\sigma^2).
\end{eqnarray}
The probability of observing a censored observation at a distance $d$ is given by
\begin{eqnarray}
P(y\geq c) = 1 - \Phi\left(\frac{c-\mathbf{x_{i}}\boldsymbol\alpha}{\sigma}\right),
\end{eqnarray}
where $\Phi$ is the cumulative distribution function (CDF) of the standard normal distribution. Now, by using $I$ as an indicator function that is set to 1 if the observation is uncensored and is set to 0 if the observation is censored, it is possible to write down the likelihood function as \cite{Tobin}

\begin{equation*}
 l(\sigma, \boldsymbol\alpha) =  \prod_{i=1}^N\left[ \frac{1}{\sigma}\phi\left(\frac{y_{i}-\mathbf{x_{i}}\boldsymbol\alpha}{\sigma} \right) \right]^{I_{i}} \left[1- \Phi\left(\frac{c-\mathbf{x_{i}}\boldsymbol\alpha}{\sigma} \right) \right]^{1-I_{i}},
\end{equation*}
where $\phi$ is the standard normal probability density function (PDF). The log-likelihood $L(\sigma,\boldsymbol\alpha)=\mathrm{ln}[l(\sigma,\boldsymbol\alpha)]$ can now be written as

\begin{equation}
\begin{aligned}
L(\sigma,\boldsymbol\alpha) = \sum_{i=1}^{N} I_{i}\left[-\mathrm{ln}\sigma+\mathrm{ln}\phi\left(\frac{y_{i}-\mathbf{x_{i}}\boldsymbol\alpha}{\sigma} \right) \right]& \\+ \sum_{i=1}^{N}(1-I_{i})\mathrm{ln}\left[ 1- \Phi\left(\frac{c-\mathbf{x_{i}}\boldsymbol\alpha}{\sigma} \right) \right]&.
\end{aligned}
\end{equation}

\noindent Using the log-likelihood, the parameters $\sigma$ and $\boldsymbol\alpha$ are estimated using
\begin{eqnarray}
[\hat{\sigma}, \hat{\boldsymbol{\alpha}}]=\underset{\sigma,\boldsymbol\alpha}{\arg\min}\{ -L(\sigma,\boldsymbol\alpha)\},
\end{eqnarray}
which is easily solved by numerical optimization of $\boldsymbol\alpha$ and $\sigma$, using for instance the method of Newton \cite{Takeshi}. In this work, we have solved this by using the fminsearch function in Matlab, which is based on a Nelder-Mead search method. The estimates obtained from OLS were used as initial values for the minimization. The presented method approach can easily be further extended, so that it supports  other pathloss models. 

\subsection{Asymptotic Variance of the ML estimator}
The asymptotic variance of the ML estimator has been derived in \cite{Takeshi} for the problem with censoring of samples where $y_{i}\leq0$. We therefore transform the data in Eq.~(\ref{model_1}), by letting
\begin{equation}
\mathbf{y}_{t}=-\mathbf{y}+c=-\mathbf{X}\boldsymbol\alpha-\boldsymbol\epsilon+c=\mathbf{X}\boldsymbol\alpha_{t}-\boldsymbol\epsilon,
\label{model_transform}
\end{equation}  
where
\begin{equation}
\boldsymbol\alpha_{t}=[-PL(d_{0})+c \ \ \ -n]^{T}.
\label{model_transform}
\end{equation}
The parameters to be estimated for the transformed data are
\begin{equation}
\boldsymbol\theta_{t}=[\boldsymbol\alpha^{T}_{t} \ \ \ \sigma^2]^{T}.
\label{model_transform2}
\end{equation}
The asymptotic variance for the ML estimates of the original parameters, $\boldsymbol\theta$, are however the same as for the parameters of the transformed data, $\boldsymbol\theta_{t}$. Therefore, we can directly use the equations found in \cite{Takeshi} to calculate the asymptotic variance as
\begin{eqnarray}
\mathrm{Avar}(\boldsymbol\theta)=\mathrm{Avar}(\boldsymbol\theta_{t})=\mathrm{diag}\left\{\left(\sum_{i=1}^{N}\mathbf{A}_{i}(\mathbf{x}_{i},\boldsymbol\theta_{t})\right)^{-1}\right\}, \label{Avar}
\end{eqnarray}
where
\begin{equation}
\mathbf{A}_{i}(\mathbf{X}_{i},\boldsymbol\theta_{t}) = 
  \begin{pmatrix} a_{i}\mathbf{x}_{i}^T\mathbf{x}_{i} & b_{i}\mathbf{x}_{i}^T \\ b_{i}\mathbf{x}_{i} & c_{i} \end{pmatrix}, 
\end{equation}
with coefficients
\begin{equation}
\begin{aligned}
a_{i} &= -\sigma^{-2}\left[z_{i}\phi_{i}-\phi_{i}^2/(1-\Phi_{i})-\Phi_{i}\right], \\
b_{i} &= \sigma^{-3}\left[z_{i}^2\phi_{i}+\phi_{i}-z_{i}\phi_{i}^2/(1-\Phi_{i})\right]/2, \\
c_{i} &= -\sigma^{-4}\left[z_{i}^3\phi_{i}+z_{i}\phi_{i}-z_{i}^2\phi_{i}^2/(1-\Phi_{i})-2\Phi_{i}\right]/4.
\end{aligned} \label{eq:coeff}
\end{equation}
Here, $\phi_{i}=\phi_{i}(z_{i})$ and $\Phi_{i}=\Phi_{i}(z_{i})$ and $z_{i}=\mathbf{x_{i}}\boldsymbol\alpha_{t}/\sigma$. In order to avoid numerical issues when calculating the coefficients in Eq.~\ref{eq:coeff}, it is worthwhile to rewrite the ratio $\phi_{i}/(1-\Phi_{i})$ as
\begin{eqnarray}
\frac{\phi_{i}(z_{i})}{1-\Phi_{i}(z_{i})}=\frac{\frac{1}{\sqrt{2\pi}}\mathrm{exp}(-z_{i}^2/2)}{1-\frac{1}{2}\mathrm{erfc}(-z_{i}/\sqrt{2})}=\frac{2}{\sqrt{2\pi}\mathrm{erfcx}(z_{i}/\sqrt{2})},
\end{eqnarray}
where $\mathrm{erfc(\cdot)}$ is the complementary error function and $\mathrm{erfcx(\cdot)}$ is the scaled complementary error function. 

As stated previously, the asymptotic variances of the parameters $\boldsymbol\theta_{t}$ are the same as for $\boldsymbol\theta$. Therefore, the asymptotic variance of the parameters $PL(d_{0})$, $n$ and $\sigma^2$ are given by the three main diagonal elements of the matrix in Eq.~(\ref{Avar}). For measurement data, an estimate of the asymptotic variance can be found by replacing the true parameter values with their estimates, $\hat{PL}(d_{0})$, $\hat{n}$ and $\hat{\sigma}^2$. Estimates of the standard errors can then be obtained simply by taking the square root of the asymptotic variance. 

The standard errors of the estimated parameters depend on many different things, such as the pathloss sample distances, the level of the censoring, the number of samples as well as the exact values of $PL(d_{0})$, $n$ and $\sigma^2$. Therefore, it is often necessary to evaluate the standard errors for each individual measurement case. Implemented Matlab codes for the ML estimator and its asymptotic variance can be found in \cite{Technical}.

\section{Results}
As an example, synthetic data at 5.6 GHz was generated according to Eq.~(\ref{eq:PowerLaw}) with known parameters ($n=2$ and $\sigma=4$) and a synthetic censoring level at $c$. The parameters were estimated using OLS and the ML method described above. The result is shown in Fig.~1. The OLS method clearly underestimates both the pathloss exponent, $\hat{n}$, as well as the standard deviation of the large scale fading, $\hat{\sigma}$. The ML method on the other hand, is able to correctly estimate both parameters in this example. Fig.~2 shows the same thing as Fig.~1, but is for measured data from a vehicle-to-vehicle (V2V) measurement campaign for NLOS scenarios at 5.6 GHz \cite{TaimoorArxiv}. In this case, the parameter estimates obtained using OLS show significantly smaller values compared to the parameter estimates for the ML method.
\begin{figure}
%
%
\includegraphics[width=1.0\columnwidth]{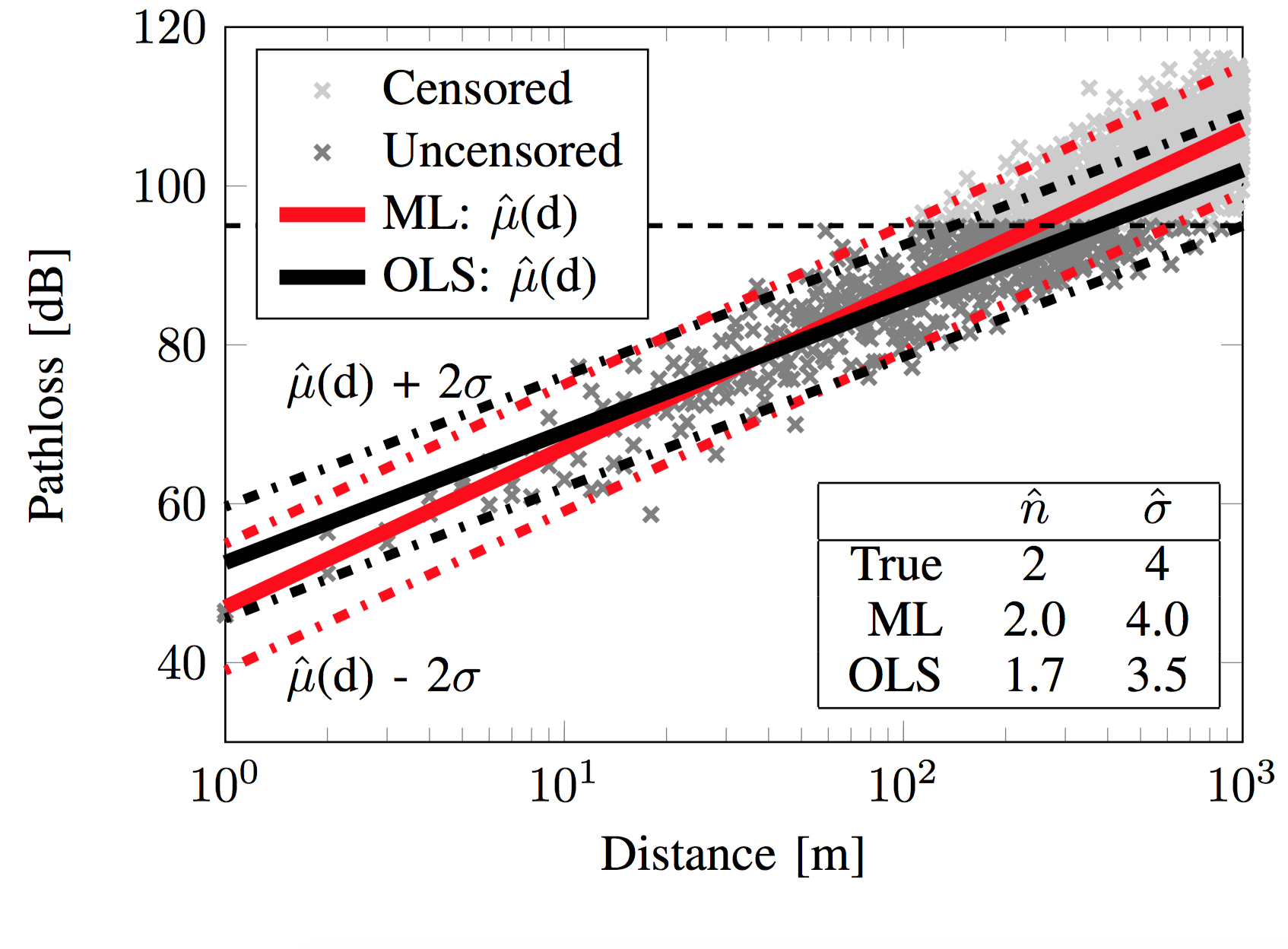}
\label{PLsynt}
\caption{Pathloss estimation based on censored synthetic data using the ML estimation method that considers censoring and using OLS without considering censoring. The ML method produces accurate estimates, whereas the OLS method underestimates $n$ and $\sigma$.}
\end{figure}
\begin{figure}
%
%
%
\includegraphics[width=1.0\columnwidth]{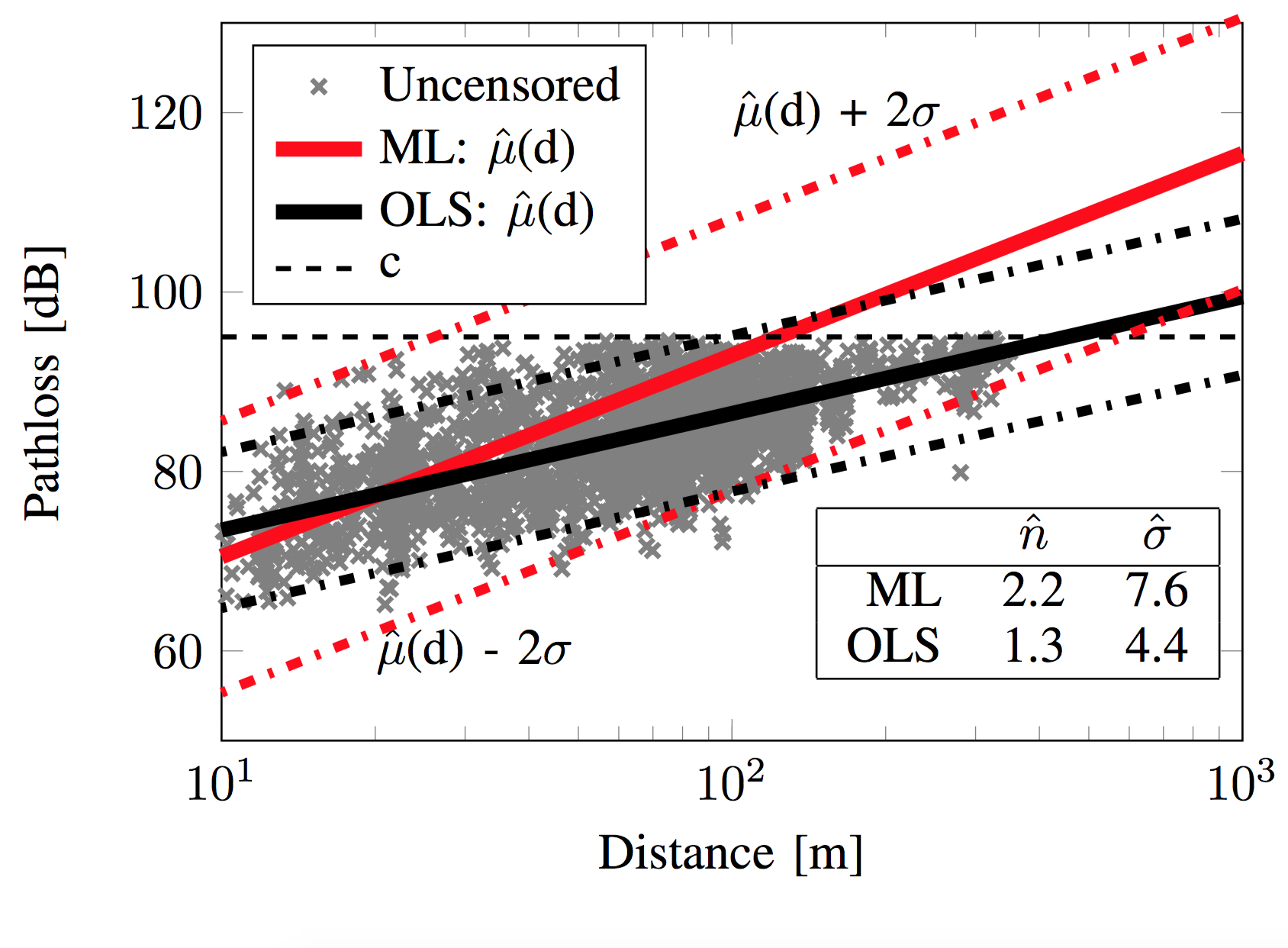}
\label{PLmeas}
\caption{Pathloss estimation of censored measurement data, using ML and OLS estimation. The estimated standard error for the ML estimates are $\mathrm{SE}(\hat{PL}(d_{0}))=0.72$ dB,  $\mathrm{SE}(\hat{n})=0.04$ and $\mathrm{SE}(\hat{\sigma}^2)=1.6$.}
\end{figure}
This large discrepancy is due to the large number of censored samples in this data set; about 45~\% of the measurement data points are censored. As a result, the OLS, which does not consider the censored samples, greatly underestimates the pathloss exponent and large scale fading. This shows the importance of taking censored samples into account when estimating the pathloss parameters.

\section{Conclusions}
In this letter, we suggest the use of a Tobit ML method \cite{Tobin} for the estimation of pathloss parameters based on censored data. When the data is censored, the standard approach of OLS, which has been widely used in the literature, is inconsistent, and yields biased estimates. The suggested ML estimator solves this problem by jointly estimating the parameters based on a censored normal distribution. Equations for the standard errors of this estimator are also provided. Using these equations, we show that the sampling distribution of the measurement samples can have a significant effect on the standard error in typical pathloss measurements. 
Using synthetic pathloss data that is censored, we also show that the ML method is able to correctly estimate the pathloss parameters, whereas OLS is biased and underestimates the pathloss exponent and the large-scale fading variance. Lastly, by using measured pathloss data from a V2V measurement campaign, we see that the ML method yields drastically different and more realistic estimates compared to the OLS method. Additional results and Matlab codes can be found in the supplementary technical report \cite{Technical}. 

\bibliographystyle{IEEEtran}
\bibliography{IEEEabrv,Pathlossbib.bib}

\end{document}